\documentclass[sigconf]{acmart}

\AtBeginDocument{%
  \providecommand\BibTeX{{%
    \normalfont B\kern-0.5em{\scshape i\kern-0.25em b}\kern-0.8em\TeX}}}

\makeatletter
\def\@ACM@checkaffil{% Only warnings
    \if@ACM@instpresent\else
    \ClassWarningNoLine{\@classname}{No institution present for an affiliation}%
    \fi
    \if@ACM@citypresent\else
    \ClassWarningNoLine{\@classname}{No city present for an affiliation}%
    \fi
    \if@ACM@countrypresent\else
        \ClassWarningNoLine{\@classname}{No country present for an affiliation}%
    \fi
}
\makeatother

\setcopyright{acmlicensed}
\copyrightyear{2024}
\acmYear{2024}
\acmDOI{XXXXXXX.XXXXXXX}

\acmConference[PASC '26]{Platform for Advanced Scientific Computing}{June 29-- July 1,
  2026}{Bern, Switzerland}
\acmBooktitle{PASC '26: Platform for Advanced Scientific Computing,
 June 29-- July 1, 2026, Bern, Switzerland} 
\acmISBN{978-1-4503-XXXX-X/18/06}

\newif\ifdraft
\drafttrue

\ifdraft
 \newcommand{\hm}[1]{{\textcolor{magenta}{ ***Heng: #1 }}}
 \newcommand{\ian}[1]{{\textcolor{red}{ ***Ian: #1 }}}
\else
 \newcommand{\hm}[1]{}
 \newcommand{\ian}[1]{}
\fi

\usepackage{tabularx}

\usepackage{color}
\usepackage[normalem]{ulem} % for strike-through

\newcommand{\rev}[1]{#1}
\newcommand{\del}[1]{}

\begin{document}

\title{Scalable Agentic Reasoning for Designing Biologics Targeting Intrinsically Disordered Proteins}


%\author{Anonymous Authors}
\author{
Matthew Sinclair$^{1,\dagger}$, 
Moeen Meigooni$^{1,\dagger}$,
Archit Vasan$^{1,\dagger}$,
Ozan Gökdemir$^{1,2}$,
Xinran Lian$^{1}$,
Heng Ma$^{1}$,
Yadu Babuji$^{2}$,
Alexander Brace$^{1,2}$,
Khalid Hossain$^{1}$,
Carlo Siebenschuh$^{1,2}$,
Thomas Brettin$^{1}$,
Kyle Chard$^{2}$,
Christopher Henry$^{1}$,
Venkatram Vishwanath$^{1}$,
Rick L. Stevens$^{1,2}$,
Ian T. Foster$^{1,2}$,
Arvind Ramanathan$^{1,2*}$,
}
\affiliation{
$^{1}$Argonne National Laboratory, 
$^{2}$University of Chicago
}


\renewcommand{\shortauthors}{Sinclair M., et al.}
%\renewcommand{\shortauthors}{Anonymous Authors}

\begin{abstract}
Intrinsically disordered proteins (IDPs) represent crucial therapeutic targets due to their significant role in disease--approximately 80\% of cancer-related proteins contain long disordered regions -- but their lack of stable secondary/tertiary structures makes them ``undruggable.''
While recent computational advances, such as diffusion models, can design high-affinity IDP binders, \del{translating these to practical drug discovery requires autonomous systems capable of reasoning across complex conformational ensembles and orchestrating diverse computational tools at scale} \rev{autonomous systems can accelerate their translation to practical drug discovery by reducing the need for expert intervention across large-scale design campaigns}.
To address this challenge, we designed and implemented StructBioReasoner, a scalable multi-agent system for designing biologics that can be used to target \del{IDPs} \rev{both IDPs and structured proteins}. 
StructBioReasoner employs a novel tournament-based reasoning framework where specialized agents compete to generate and refine therapeutic hypotheses, naturally distributing computational load for efficient exploration of the vast design space. 
Agents integrate domain knowledge with access to literature synthesis, AI-structure prediction, molecular simulations, and stability analysis, coordinating their execution on HPC infrastructure via an extensible federated agentic middleware, Academy. 
We benchmark StructBioReasoner across Der f 21 and NMNAT-2 and demonstrate that over 50\% of 787 designed and validated candidates for Der f 21 outperformed the human-designed reference binders from literature, in terms of improved \rev{in silico} binding free energy. 
For the more challenging NMNAT-2 protein, we identified three binding modes from 97,066 binders, including the well-studied NMNAT2:p53 interface. 
Thus, StructBioReasoner lays the groundwork for agentic reasoning systems for IDP therapeutic discovery on Exascale platforms.
\end{abstract}

%Let's say something about Academy and orchestration across labs.
% 250 words max
% Apoarently, 1/3 of the human proteome is IDPs (I cite this in the paper). We can say something about that if you want. Kind of interesting, might capture interest. We don't know how to drug this stuff, even though it's so common in diseases.





\maketitle

\section{Introduction}
Intrinsically disordered proteins (IDPs) comprise between 30\% and 40\% of the human proteome and play critical roles in cellular regulation, signaling, and disease pathogenesis~\cite{holehouse2024idp}. Approximately 80\% of cancer-associated proteins contain long intrinsically disordered regions, making these proteins essential therapeutic targets~\cite{moellering2024undruggable}. Yet IDPs have historically been classified as ``undruggable'' because they lack the stable three-dimensional structures and well-defined binding pockets that conventional drug discovery methods require. The conformational heterogeneity of IDPs—their existence as dynamic ensembles rather than fixed structures—poses fundamental challenges for rational biologics design~\cite{wang2023idp}.

Recent computational breakthroughs have begun to address this challenge. Methods based on diffusion models, including modified RFdiffusion and the Logos framework have demonstrated the ability to generate high-affinity protein binders (K$_d$ = 3--100 nM) for previously intractable IDP targets~\cite{liu2025rfdiffusion,wu2025logos}. These approaches achieve success by freely sampling conformations of both target and binder during the design process, effectively navigating the conformational landscape that makes IDPs difficult to target. However, translating these individual computational capabilities into practical drug discovery requires autonomous systems that can: (1) reason about which computational strategies to apply for a given IDP target, (2) orchestrate diverse tools including structure prediction, molecular dynamics, and thermodynamic analysis, and (3) scale efficiently across high-performance computing resources to explore vast design spaces.

This challenge sits squarely at the intersection of artificial intelligence (AI), structural biology, and high-performance computing (HPC). While AI agents have shown promise for autonomous scientific discovery in biology, chemistry and materials science~\cite{boiko2023coscientist,bran2024chemcrow}, existing agentic systems for protein engineering have focused primarily on ordered proteins or small molecule design. No current framework specifically addresses the unique requirements of IDP-targeting biologics: reasoning about conformational ensembles, selecting appropriate computational tools from the expanding arsenal of IDP-specific methods, and executing compute-intensive workflows at HPC scale. Furthermore, the computational demands of IDP biologics design—involving ensemble generation, exhaustive conformational sampling, and iterative refinement—necessitate scalable architectures that can leverage distributed GPU resources and minimize tool-calling overhead.

\begin{figure*}
    \centering
    \includegraphics[width=\linewidth]{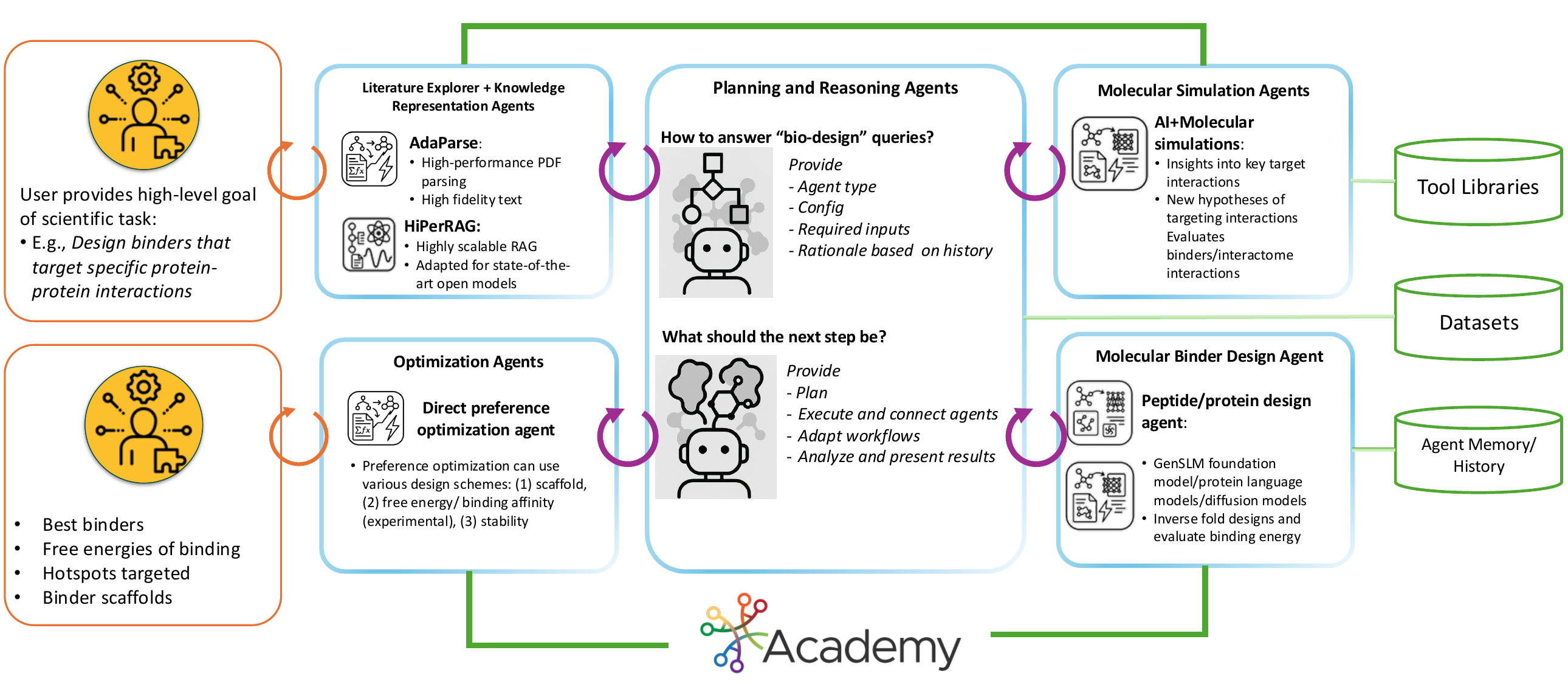}
    \caption{StructBioReasoner design/architecture. 
A user provides a high-level design goal, and the agent system dynamically selects from  specialized agents to execute the task iteratively. Each specialized agent is orchestrated via the Academy agentic framework and has access to a variety of tools, datasets including literature and other related data, as well as history of actions taken by the agents to run the tools. The planner and reasoner agents act in tandem to present results to the user, which can then be refined in subsequent interactions.}
%(B) Case study showing the end-to-end workflow for binder design  targeting NMNAT-2 integrating literature mining, structural prediction, molecular design, experimental validation, and optimization techniques.}
    \label{fig:sbr_arch}
\end{figure*}

We present StructBioReasoner, a scalable multi-agent system for autonomous IDP-targeting biologics design that addresses these challenges. 
\del{This system (\autoref{fig:sbr_arch}) employs a tournament-based reasoning framework where specialized agents--focused on structural stability, evolutionary conservation, energetic optimization, and rational design principles--compete to generate and refine hypotheses for engineering peptides/biologics targeting IDPs \rev{and IDRs}. The tournament architecture enables parallel hypothesis evaluation and ranking, naturally distributing computational work across available resources. Each agent integrates domain knowledge with access to computational tools including various AI-structure prediction approaches, molecular dynamics (MD) simulation, and physics-based stability prediction. The system reasons about which tools to invoke for a given design challenge and coordinates their execution on HPC infrastructure through workflow orchestration frameworks.}
\rev{This system (\autoref{fig:sbr_arch}) employs a tournament-based reasoning framework in which specialized agents iteratively generate, evaluate, and select among competing hypotheses spanning target hotspots, design strategies, and binder scaffolds for engineering peptides/biologics targeting IDPs and IDRs. Each iteration of the design loop constitutes a tournament round, with successful strategies promoted into further refinement and failing directions annotated and culled. Agents integrate domain knowledge with access to computational tools including AI-structure prediction, molecular dynamics (MD) simulation, and physics-based stability prediction, reasoning about which tools to invoke and coordinating their execution on HPC infrastructure through workflow orchestration frameworks.}

Our contributions are threefold. First, we demonstrate a multi-agent reasoning architecture designed to address the conformational complexity of IDP targets, where agents must reason about ensemble properties rather than a single structure. 
Second, we present a scalable tournament framework that enables efficient parallel hypothesis generation and evaluation, addressing the computational intensity of IDP biologics design. 
Third, we validate our approach through case studies on two model
\del{IDP systems, including (1) house dust mite allergen Der f 21 and (2) Nicotinamide mononucleotide adenylyltransferase 2 (NMNAT-2) showing that our agentic reasoning framework can identify} 
\rev{systems of increasing disorder: (1) house dust mite allergen Der f 21, a structured protein used to benchmark our framework, and (2) Nicotinamide mononucleotide adenylyltransferase 2 (NMNAT2), a protein containing significant disordered regions, for which our agentic reasoning framework designs binders targeting both ordered and disordered domains while identifying}
stabilizing mutations matching literature-validated strategies while effectively utilizing distributed supercomputing resources.
This work establishes a foundation for autonomous discovery of IDP-targeting therapeutics on emerging exascale platforms, advancing both the scientific understanding of these challenging targets and the computational methods required to drug them effectively.
\rev{The StructBioReasoner workflow, including all source code, prompt templates, and configuration files, is publicly available at \url{https://github.com/IDeA-ANL-ORNL/StructBioReasoner/}}

\section{Related Work}

We review related work in agentic systems for biologics design, computational methods for IDPs, and HPC for biological AI.

\subsection{Agentic Systems for Biologics Design}

The integration of large language models with specialized tools has enabled autonomous scientific discovery across domains. In chemistry and biology, platforms such as ChemCrow, Kosmos, and CoScientist demonstrated that LLMs could plan syntheses, control laboratory robotics, and iteratively optimize reactions~\cite{bran2024chemcrow,mitchener2025kosmosaiscientistautonomous,boiko2023coscientist}. Together with the rapid maturation of open-source frameworks \cite{langchain2023,langgraph2024,crewai2024} for building agentic workflows—enabling explicit control flow, tool use, and multi-agent coordination—these successes demonstrate the feasibility of agentic scientific workflows for orchestrating complex, multi-stage experimental campaigns.

Protein engineering further leverages these capabilities. Virtual Lab coordinates multiple specialized agents under the guidance of an LLM Principal Investigator to execute therapeutic protein design, producing 92 nanobody designs with over 90\% expression success~\cite{swanson2025virtuallab}. ProtAgents and SAMPLE extended this approach by integrating LLM reasoning with physics-based simulations, structural analysis, and cloud-executed closed-loop optimization~\cite{ghafarollahi2024protagents,rapp2024sample}. More comprehensive platforms like DrugAgent and PharmAgents span full pharmaceutical pipelines~\cite{liu2024drugagent}, yet these systems primarily focus on ordered proteins or small molecules and do not address the structural heterogeneity of intrinsically disordered proteins (IDPs).

StructBioReasoner advances this by introducing tournament-based multi-agent reasoning tailored to IDPs.
Critically, it designs biologics autonomously to disrupt protein-protein interaction networks in cellular pathways, integrating retrieval-augmented generation, conformational ensemble modeling, structural prediction, and multi-agent decision-making to target biologically relevant interactions rather than isolated protein structures.

\subsection{Computational Methods for IDPs}

Intrinsically disordered proteins (IDPs) \rev{and proteins with intrinsically disordered regions (IDRs)} present two computational challenges: capturing their conformational heterogeneity and designing binders capable of engaging highly flexible targets. Although recent advances have substantially improved both structure prediction and binder engineering, workflows are still fragmented and demand expert intervention.
Accurate structure prediction of IDPs remains a central challenge due to their intrinsic flexibility and lack of stable tertiary structure.
AlphaFold~3 introduced disorder-aware training through cross-distillation, significantly improving predictions for disordered regions \cite{abramson2024af3}.
Nevertheless, evaluation on DisProt benchmarks revealed that $\sim$22\% of predictions incorrectly impose order on experimentally validated disordered regions \cite{gopalan2025af3eval}, highlighting the persistent difficulty of modeling IDP ensembles.
Complementary tools such as AlphaFold-Multimer, which can model IDP–partner complexes \cite{omidi2024alphafold}, and specialized frameworks like SPARROW and DisoFLAG \cite{lotthammer2024sparrow}, provide additional IDP-specific prediction capabilities.
More recent developments, including diffusion-based approaches like IDPFold \cite{zhu2024idpfold} and maximum-entropy reweighting methods that integrate molecular dynamics with experimental data \cite{borthakur2025ensemble}, enable principled generation of conformational ensembles rather than single-structure predictions.

In parallel, substantial progress in IDP-targeted binder design has demonstrated that high-affinity recognition of disordered proteins is computationally feasible.
A modified RFdiffusion model trained on two-chain complexes enables joint sampling of both binder and IDP conformations, producing binders with K$_d$ values as low as 3–100 nM and achieving functional inhibition of amyloidogenic targets \cite{liu2025rfdiffusion}.
Concurrently, the Logos framework employs a curated library of 1000 prefabricated binding pockets to design tight binders for 39 of 43 structurally diverse targets \cite{wu2025logos}.
Despite their strong performance, these systems remain isolated tools that require manual selection, coordination, and interpretation.

StructBioReasoner builds directly on these methodological advances by integrating structure prediction, ensemble generation, and binder design into an autonomous, multi-step reasoning framework. 
By determining when ensemble modeling is required, selecting the appropriate computational tools, and iteratively interpreting and refining intermediate outputs, it unifies capabilities that have previously operated in isolation and provides a coherent path toward end-to-end computational design for IDPs.

\subsection{Scalable HPC for Biological AI}

Scaling protein structure prediction and design to HPC resources has progressed significantly. 
%ScaleFold reduced AlphaFold pretraining time from 7 days to 10 hours using 2080 H100 GPUs through non-blocking data pipelines and custom CUDA kernels~\cite{zhu2024scalefold}.
NVIDIA's BioNeMo Framework established industry standards for biological AI at scale, achieving 59.2\% model FLOPs utilization and near-linear scaling to 256 GPUs~\cite{john2024bionemo}. OpenFold provided the open-source trainable AlphaFold with custom CUDA attention kernels enabling inference on sequences exceeding 4000 residues~\cite{ahdritz2024openfold}. For therapeutic discovery/design workflows, platforms like aweVS demonstrated screening of 1.6 billion compounds in under one day using 26,000 GPUs~\cite{10596529}, while NVIDIA NIM blueprints achieved 10$\times$ acceleration for virtual screening~\cite{nvidiavs2024} and the Simple SMILES Transformer workflow was deployed across 256 compute nodes of ALCF's Aurora supercomputer to screen 22 billion compounds in 40 minutes~\cite{10596529}. 
%Workflow orchestration systems including Parsl, FuncX, and Colmena have enabled Python-based parallel workflows scaling from laptops to supercomputers~\cite{babuji2019parsl,chard2020funcx,ward2021colmena}. Recent integration of RADICAL-Pilot with Parsl demonstrated sustained throughput exceeding 1500 tasks/second on Frontier~\cite{yokelson2024radicalparsl}.
Most relevant to our work, the MProt-DPO framework achieved greater than 1 ExaFLOPS sustained performance for multimodal protein design workflows across five supercomputers through direct preference optimization~\cite{dharuman2024mprot}, and RL-based biophysical scoring \cite{rlforbiophysicalscoringdharuman}.
Additionally, the IDEAL project at Argonne explicitly targets scalable agentic workflows on the ALCF Aurora supercomputer for IDP-targeting anticancer therapeutics~\cite{ideal2024}.

%StructBioReasoner complements these efforts for IDP biologics discovery by providing an agentic reasoning layer that (1) makes autonomous decisions about tool selection, parameter configuration, and computational resource allocation and (2) orchestrates  HPC-scale tools across federated HPC resources via the Academy agentic framework~\cite{pauloski2025academy}.
\section{StructBioReasoner Architecture} %Design and Implementation}
%StructBioReasoner enables scalable biologics design to target protein-protein interfaces, with a focus on IDPs (Fig~\ref{fig:sbr_arch}). 
StructBioReasoner enables scalable biologics discovery by providing an agentic reasoning layer that makes autonomous decisions about tool selection, parameter configuration, and computational resource allocation and orchestrates HPC-scale tools across federated HPC resources via the Academy agentic framework~\cite{pauloski2025academy}: \autoref{fig:sbr_arch}.
Our approach includes several core agents, each given access to a subset of literature synthesis, knowledge representation, molecular design, and reasoning tools to accomplish its expert task. 
\rev{
The framework is designed to be modular and extensible: each agent defines a standard interface over its inputs and outputs, allowing new tools to be integrated as drop-in replacements for any subtask. 
For instance, the binder design agent currently supports diffusion-based, iterative refinement, and language model-driven strategies, but its interface readily accommodates emerging generative approaches such as flow matching models as they mature. 
Likewise, the reasoning and planning agents are LLM-agnostic; newer reasoning models can be substituted without modifying the agent coordination logic. 
Table 1 provides a snapshot of the current tool--agent mappings and is expected to grow as new methods become available.
}

For a given target, our \textit{Reasoning Agent} first engages a RAG-based literature agent, HiPerRAG~\cite{gokdemir2025hiperrag} to infer protein-protein interactions (PPIs), typically referred to as the interactome.
The resulting PPIs and the entire interactome are examined by the Reasoning Agent to identify pathways of interest to intervene via targeted design of peptide-based therapeutics  before selecting a subset of the interactome to the protein structure prediction agent to predict the putative complex structures.
The \textit{Molecular Simulation Agent} then builds and executes long-timescale MD simulations on each interactome pair, and returns a confidence score for each pair representing the most probable interface to target with biologics. This allows the Reasoning agent to determine the most likely interface from the target protein's perspective (an IDP loop, surface or any adjacent region) that can be used to design the peptide/protein-based binder. 

\rev{
\begin{figure*}[th!]
\begin{center}
%\footnotesize\begin{tabularx}{0.935\textwidth}{ >{\color{blue}}c | >{\color{blue}}c | >{\color{blue}}c | >{\color{blue}}c | >{\color{blue}}c } % comment this guy out after revisions
\footnotesize\begin{tabularx}{0.935\textwidth}{ c | c | c | c | c } % uncomment this after revisions are done MM
 \textbf{Task} & \textbf{Agent} & \textbf{Tools} & \textbf{Input} & \textbf{Output} \\ 
 \hline 
    RAG & HiPerRag Agent & HiPerRag &  User query, literature store & Response \\  
    Reasoning & Planning and Reasoning Agent & GPT-OSS-120B & History of tool calls, configs & Next tool call, config \\
    Folding & Protein Structure Prediction Agent & Chai-1, Boltz-2x & Protein sequence & Protein structure \\
    Design & Binder Design Agent & Bindcraft, Chai, ProteinMPNN & Protein binder scaffold & Improved protein binder sequence, structure \\
    & Chroma Agent & Chroma & Protein target structure & Protein binder sequence, structure \\
    Simulation & Molecular Simulation Agent & OpenMM, molecular-simulations & Protein structure & Protein trajectory \\
    Analysis & Free Energy Agent & MM-PBSA & Protein trajectory & Binding free energy \\
    & Simulation Analysis Agent & Simple MD analyses & Protein trajectory & RMSD, RMSF, SASA, RoG \\
\label{table:agents}
\end{tabularx}
\caption*{Table 1. Summary of StructBioReasoner agents and their computational roles. Each agent is specialized for a distinct task in the biologics design pipeline, with access to specific tools that process defined inputs and produce structured outputs for downstream agents.}

\end{center}
\end{figure*}
}

Following this, the Reasoning agent decides on a binder design strategy from several state-of-the-art approaches including: conditioned diffusion \rev{(} \del{or flow modeling (e.g. RFDiffusion~\cite{watson2023novo}, }
Chroma~\cite{ingraham2023illuminating} 
\del{and SimpleFold~\cite{wang2025simplefold}}), iterative refinement of target scaffolds 
(\del{e.g.} BindCraft~\cite{pacesa2025one}), or sequence design utilizing protein embeddings in protein or genome-scale language models (\del{e.g.}xTrimo-PGLM~\cite{chen2025xtrimopglm} or GenSLM~\cite{zvyagin2023genslms}).
When the Reasoning agent is satisfied with a set of binders, they are further assessed by the MD agent using molecular mechanics Poisson-Boltzmann surface area (MM-PBSA)- based approximation for free energy calculations~\cite{miller2012mmpbsa}. The MM-PBSA approach constitutes a reasonable trade-off between the computational costs versus accuracy for estimating binding free energies. 
After each round of molecular design, the Reasoning agent evaluates the quality of designed binders, deciding either to commit to further rounds of development using the chosen methodology, to switch design methods, or to continue probing the existing suite of binders through advanced simulation.

After designing a large set of binders, subsequent rounds of design can begin by first employing reinforcement learning using direct preference optimization on all quality designed sequences.
Each sequence is binned into preferred/unpreferred pairs by utilizing a custom score comprising \del{of} several structure- and sequence-based metrics including binding free energy, root mean square deviation (RMSD)/root mean square fluctuation (RMSF) for molecular stability, and developability (net charge, hydrophobicity, etc.).
This iterative approach allows us to leverage the rich embedding space from the language model of choice within the context of the design experiment.

\subsection{\del{Planning and }Reasoning Agent}
The Reasoning \del{and Planning} agent is an LLM-driven controller for the entire binder-design workflow.
Built in spirit by integrating the techniques from the Google Co-scientist and Virtual Lab~\cite{boiko2023coscientist, swanson2025virtuallab}, the Reasoning agent performs adaptive scientific reasoning: interpreting new results, updating strategy, and coordinating the next design or evaluation step.

Whenever a specialized agent (HiPerRAG, protein structure prediction, MD, Analysis, Optimization) produces new outputs, the Reasoning agent compiles a structured state summary.
This summary includes the latest results, the research goal, target-protein information, and recent decisions.
The Reasoning agent then analyzes the state and recommends the next action (e.g., generate new candidates, run simulations, compute binding energetics, or terminate the workflow), along with a short rationale and a confidence estimate.
The Reasoning agent then (1) converts this recommendation into updated parameters for the next iteration, using the full historical record of decisions and outcomes to maintain continuity and (2) executes and connects agents.
\del{
The Reasoning agents are model-agnostic and can route reasoning requests to public LLMs such as GPT-OSS-120B~\cite{agarwal2025gpt} or to biology-specialized reasoners like BioMNI~\cite{huang2025biomni}, depending on the task.
}
\rev{
The Reasoning agent uses GPT-OSS-120B~\cite{agarwal2025gpt}, an open-source model hosted and served on ALCF Sophia, for inference with a sampling temperature of 0.7 and a maximum output length of 32768 tokens to generate structured JSON outputs. While the architecture is model-agnostic and can route reasoning requests to other LLMs depending on the task, we chose GPT-OSS-120B for its open-source availability and strong%ish
reasoning capabilities.
}
%General models handle broad planning and heuristics, while domain-tuned models interpret structural, biochemical, or sequence-level details more reliably. This hybrid approach improves both robustness and biological accuracy.

By continually integrating new data and refining its decision policy, the system forms a closed-loop controller that autonomously explores hypotheses, validates predictions, scores binding performance, and determines when optimization has converged.
\rev{This closed-loop design underpins the tournament-based framework: each iteration is a round in which the Reasoning Agent proposes competing strategies such as target hotspot, design methodology, and binder scaffold seeds while specialized agents enhance and evaluate these hypotheses.  
Outcomes are scored using task-specific metrics including binding free energy (MM-PBSA), structural stability (RMSD/RMSF), and interaction energetics, which are aggregated in a structured history accessible to the Reasoning Agent.
Successful strategies are promoted in further rounds of refinement and failing directions are culled, distilling the Reasoning agent with design patterns and target-specific intuition without model retraining.}
%Hypothesis generation is largely LLM-driven, and preliminary success is driven by choice of the Reasoning agent.
%Herein, we have built Jnana to manage Reasoning agents to drive this integration. 
%Any LLM with a public-facing API, including specialized models finetuned on biological data such as Biomni~\cite{}.
%We also show that self-hosted open models such as GPT-OSS-120B are performant, allowing this approach to be run fully self-contained.
%For a given topic such as designing binders to disrupt the cancer target NMNAT-2, the Jnana system first generates a set of 3-5 expert subagents which engage in a scientific debate which includes generation and reflection capabilities.
%Over many rounds of generation a number of hypotheses are generated and ranked in a tournament setting and assigned an ELO score such that the winning hypotheses can be understood both by humans and other agents.
%
%Initial hypotheses generated through multi-agent debate represent informed but untested conjectures about target biology and design strategies. 
%These nascent hypotheses are then refined to generate actionable, computationally validated design campaigns through three interconnected mechanisms: literature-grounded augmentation, structural validation, and iterative feedback integration.
\subsubsection{Cross-hypothesis learning}
Beyond refining individual hypotheses, the system also distills generalizable patterns across the full hypothesis population.
When multiple hypotheses address related protein families or interaction motifs, the Reasoning agent identifies shared success and failure modes.
For example, if hydrophobic hotspots repeatedly outperform charged residues for a given target class, this insight is promoted into a soft design constraint for future hypothesis generation.

This cross-hypothesis learning is supported by a structured memory system that records decisions, research plans, results, and reasoning outputs.
Analogous to human cognition, the system maintains both long-term and short-term memory. Long term memory represents persistent items the reasoner should not discard (e.g., key hotspots, top binders at each stage, experimental data) while short-term memory is the evolving sequence of workflow decisions and task outcomes.
To effectively manage context length and maintain strategic continuity, we periodically trim the short-term memory while retaining the long-term memory in its entirety, updating it as new information becomes available. Across extended design campaigns, this mechanism allows the system to accumulate target-class-specific design intuition without requiring explicit model retraining.

\subsubsection{Human-in-the-loop checkpoints}
While StructBioReasoner operates autonomously, the refinement process exposes optional checkpoints for human expert review. 
At configurable intervals or when hypothesis scores plateau, the system generates a structured summary of top-ranked hypotheses, supporting evidence, and remaining uncertainties.
Domain experts can inject new hypotheses, adjust scoring weights, or override agent decisions. 
These interventions are logged and incorporated into the knowledge graph, ensuring that human expertise augments rather than disrupts the autonomous workflow.

\subsection{HiPerRAG Agent}

\begin{figure*}
    \centering
    \includegraphics[width=0.95\linewidth]{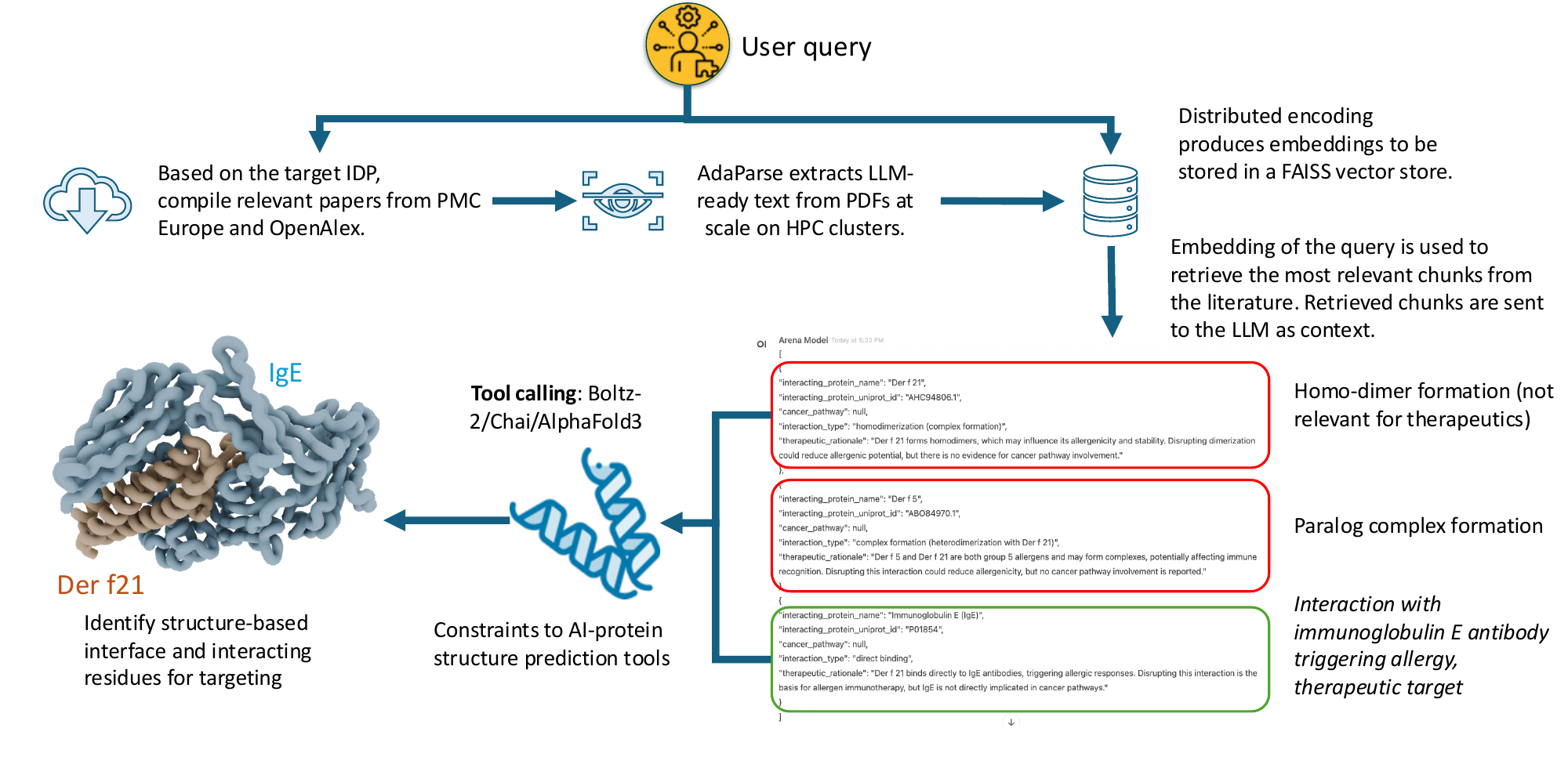}
    \caption{HiPerRAG agent inferred PPIs for Der f21 protein. A precomputed vector store comprising scientific articles from bioRxiv, arXiv and select journals is queried to augment Reasoning agent hypotheses. This generates a list of interactions that are then annotated with specific residue-level interactions that mediate Der f 21 PPIs. These can then be automatically input to protein structure prediction agents to infer the co-folded structure.}
    \label{fig:hiperrag}
\end{figure*}

%Approximately $\sim$32.8\% of the human proteome consists of intrinsically disordered residues \cite{alderson2023}, yet intrinsically disordered proteins (IDPs) remain understudied relative to structured targets such as kinases and GPCRs. 
Frontier LLMs are weakly informed about IDP biology, leading to biased reasoning and hallucinations in IDP-related tasks.
To overcome this limitation, we use retrieval-augmented generation (RAG) \cite{lewis2020retrieval}, which augments an LLM at inference time with expert-curated, literature-grounded knowledge about IDP biology, which may be target specific (in addition to the LLM's parametric memory: \autoref{fig:hiperrag}). We constructed a corpus of 1520 full-text papers on NMNAT-2 and about 38 full-text papers on Der f 21 using an automated pipeline integrating Europe PMC \cite{rosonovski2024europe}, OpenAlex \cite{priem2022openalex}, Crossref \cite{hendricks2020crossref}, and Unpaywall \cite{dhakal2019unpaywall}, with de-duplication by DOI/PMCID and OA-prioritized PDF recovery.

PDFs are parsed with AdaParse \cite{siebenschuh2025adaparseadaptiveparallelpdf} and segmented into semantically coherent chunks using a similarity-based Semantic Chunking algorithm \cite{hearst1997text,utiyama2001statistical}. Four-sentence buffers are embedded with an appropriate encoder of choice (in this case, PubMedBERT \cite{gu2021domain}) and chunk boundaries are defined at large embedding-distance discontinuities (95th percentile). Chunk embeddings are generated with SFR-Embedding-Mistral \cite{SFRAIResearch2024}, L2 normalized \cite{van2017l2}, and indexed in a FAISS \cite{johnson2019billion} vector store. Implemented via HiPerRAG \cite{gokdemir2025hiperrag}, this retrieval layer dynamically supplies relevant evidence to an arbitrary LLM to generate responses at inference time (\autoref{fig:hiperrag}). Retrieved evidence is then distilled into structured assertions (binding interfaces, affinities, post-translational modifications, cancer/ disease-relevant mutations) and assembled into a hypothesis-specific knowledge graph shared across agents. Conflicting findings and unresolved gaps in the literature are explicitly surfaced as uncertainties, reducing hallucinations and improving scientific grounding.

\subsection{Protein Structure Prediction Agent}
Literature-augmented hypotheses undergo structural validation to assess feasibility before committing to expensive design campaigns. 
For each proposed target interface, the system executes a validation cascade comprised of structural analysis and molecular dynamics simulation.
The system queries the PDB for experimental structures of the target and its interaction partners if available. 
When experimental structures are unavailable, the structure prediction tool dispatches Chai-1~\cite{boitreaud2024chai} or Boltz-2x~\cite{passaro2025boltz} predictions, with disorder-aware interpretation of confidence scores. 
The use of such models provides the additional benefit of enabling the knowledge graph to inject experimentally defined constraints into co-folding, such as key residues participating in complex formation.

\subsection{Molecular Simulation Agent}
The molecular simulation agent employs \rev{the} MD backend in an \del{in-house} \rev{internal} Python library, using the OpenMM engine~\cite{eastman2023openmm} and \rev{the} AMBER 19 forcefield~\cite{tian2019ff19sb}. 
With simply the path to an input PDB file, the MD agent generates a fully solvated and neutralized system using explicit solvent for detailed simulations, or an implicit solvent system for faster sampling, and deploys the simulation at scale.
\rev{Simulation parameters such as temperature (310 K), integrator (LangevinMiddleIntegrator), and timestep (4 fs, enabled by hydrogen mass repartitioning at 1.5 amu per hydrogen) are set by default, while the Reasoning agent makes higher-level decisions: whether to use implicit or explicit solvent to modulate sampling resolution and efficiency, and how long the simulation should run.}

For predicted or experimental complexes, the MD agent automatically performs rapid equilibration (10-50 ns based on the size of the complex) to assess interface stability.
The dynamics information from the simulations, such as interfaces that dissociate during equilibration or high root-mean-squared-fluctuation (RMSF) at proposed hotspot residues, is passed to the Reasoning agent, which then prioritizes or deprioritizes these candidates accordingly. 
Stable interfaces can be further clustered by all-to-all contact maps and characterized by interaction energies measured from simulation, providing the Reasoning agent with a comprehensive understanding of the various binding sites probed by interactome-scale simulation datasets.

Hypotheses that fail structural validation at this stage are not discarded, but rather annotated with failure modes and returned to the Reasoning agent. 
This feedback enables hypothesis revision; for example, shifting focus from a flexible loop region to an adjacent structured domain, or reconsidering the interaction partner.
If it is deemed that the existing sampling of a given target is insufficient, further rounds of MD simulation can be performed before other tasks are allowed to proceed.

\subsection{Binder Design Agent}
To perform the task of binder design, we have implemented several complementary design strategies into various binder design agents.
The core of our design strategy is inspired by recent efforts in biologics design, which include diffusion with inverse folding and alternating folding-refinement rounds via inverse folding~\cite{dauparas2022robust,ingraham2023illuminating,boitreaud2024chai,passaro2025boltz,watson2023novo}.
Here, we wrap each workflow into agents that, when invoked, are able to ingest decisions from the Reasoning agent to guide binder design, including hotspotting specific regions of the target, the initial binder fold class, post-translational modifications and other features that a human researcher would typically choose to seed rounds of design.

The design agents deliver concise, high-level summaries of their experimental results, allowing the Reasoning agent to  understand the success rate and viability of different experimental types for a specific target. 
For example, the reasoner might strategically deploy multiple BindCraft agents, assigning each to design a different binder modality (e.g., an affibody, an affitin, or a nanobody). 
After reviewing the experimental outcomes, successful designs are advanced, and unsuccessful ones are culled in subsequent optimization rounds.
This enables the agentic framework to leverage the advantages of multiple design philosophies while mitigating failure modes through broad coverage of the design space in terms of both structure-based and sequence-based design.

\vspace{-7pt}

\subsection{Analysis Agents}
Throughout the framework, there is a need to interject many types of custom analyses to assess task progression.
To bypass the limitations of relying solely on LLM contextual interpretation and exhaustive code execution, we employ a set of curated, robust analysis agents. 
These specialized agents handle the routine analytical tasks that parallel human expert workflow.
This abstraction of analysis into a custom agent enables different parallelism schemes that are optimized for these often CPU-bound tasks in a predominantly GPU-driven framework.

One such analysis agent, dubbed the Free Energy Agent, performs calculations using MM-PBSA~\cite{case2023ambertools}.
MM-PBSA balances the rigor of free energy calculations with the speed of several approximations, making it an appropriate choice for the high-throughput nature of our framework.
The Free Energy Agent utilizes an in-house Python-based workflow, which is both simplified in execution and more easily lends itself to parallel execution in our framework~\cite{molecular_simulations}.
For any single simulation or set of simulations for which the Reasoner would like free energy data, the Free Energy Agent performs all necessary pre- and post-processing of existing trajectory data and returns the mean and standard deviation of the free energy of binding between two groups of atoms.
This type of analysis enables more robust ranking of generated binders and provides another data point for reimagining the growing embedding space of a given swath of peptide binders.

Another analysis agent, the Simulation Analysis agent, performs standard MD simulation analyses, primarily utilizing the MDAnalysis~\cite{gowers2019mdanalysis} and mdtraj~\cite{mcgibbon2015mdtraj} Python libraries.
These include simpler metrics such as RMSD, RMSF, solvent-accessible surface area (SASA) and radius of gyration (RoG), but also more complex calculations like interaction energies.
Interaction energies are computed using the OpenMM MD engine~\cite{eastman2023openmm}, and are reported both in terms of all-to-all energies and residue-wise interaction energies, wherein we generate residue footprints.
These energy footprints establish a foundation to experimentally verify results from the RAG agent.
By demonstrating that interfaces or otherwise important residues from literature are also found to be highly interacting \textit{in silico}, we improve the confidence of our experiments in the eyes of the Reasoning agent.

\vspace{-7pt}

\subsection{Optimization Agents}
Additionally, the framework incorporates an optimization agent to guide the generative process toward biologically effective binders.
The optimization agent systematically compares candidate designs using structural metrics, literature-informed interface features, and experimental validations.
These comparisons generate preference pairs that directly fine-tune the generative policy toward more viable binders using the direct preference optimization approach~\cite{rafailov2023direct}.

During each design cycle, the optimization agent consumes outputs from upstream analyses, identifies consistent preference signals, and updates the model to favor designs that better meet functional goals. 
The result is an iterative, evidence-driven feedback loop that steadily improves binder quality without full retraining.

\vspace{-3pt}

\subsection{Tool Development}
% All agents are implemented using Academy~\cite{pauloski2025academy}, which provides a unified execution layer for running agentic workflows across heterogeneous and federated computing.
% Academy allows agen
% All tools are wrapped as agents using Academy~\cite{pauloski2025academy} to ensure they are easily deployed and scaled in any HPC environment.
% This definition covers how agents are initialized and shut down gracefully, a capability essential for leveraging Parsl \cite{babuji2019parsl}, which, in turn, allows us to scale a given task across thousands of computational nodes. 
% Each agent is provisioned with the necessary tools to execute its task, along with specific instructions on how to contribute its results meaningfully to the evolving hypothesis. This ensures that the orchestrating core Reasoning agent receives clear, actionable data.
% By exposing tools in such configurations, we are able to extend beyond a typical workflow-style approach, allowing the Reasoning agent to ingest information as it becomes available and make on-the-fly decisions about where to spend more or less time enriching the running hypothesis.
We build agents on Academy~\cite{pauloski2025academy}, a unified execution layer for coordinating agentic workflows across heterogeneous and federated computing resources. 
Academy is designed to separate high-level decision logic from execution mechanics, enabling agents to run on HPC systems.
%At the core of each agent is an Academy Manager, which serves as the execution controller responsible for dispatching tasks, managing dependencies, and interfacing with external compute resources. 
Academy integrates natively with Parsl \cite{babuji2019parsl} for workflow execution, allowing agents to express computational tasks in a portable manner while leveraging system-specific configurations for scheduling, data movement, and fault tolerance. 
This enables the same agent definition to scale from small exploratory runs to large, massively parallel workloads without modification.
%Academy leverages Globus Compute \cite{foster2011globus, allen2012software} to allow agents to submit and monitor tasks on remote endpoints in federated settings.
%This capability is particularly important for our workflow to span across HPC centers with diverse computational infrastructure at multiple national laboratories.
Academy leverages Globus Compute~\cite{foster2011globus, allen2012software}  to let agents submit and monitor tasks on remote endpoints in federated environments. This capability allows workflows to run across HPC centers with diverse infrastructure at multiple national laboratories.
%Academy abstracts these details behind a consistent execution interface, allowing agents to reason about what to run while the framework determines where and how it is executed.

%\subsubsection{Academy agent architecture}
%Each agent is given a section of the main configuration YAML and a system-appropriate Parsl configuration.
%The core components of an agent include an Academy Manager object~\cite{pauloski2025academy} and methods for both generating its own hypothesis, and integrating this hypothesis into the running input hypothesis via a user-defined Analysis dataclass.
%The Manager orchestrates tool calling in an expressive but consistent manner such that it can be scaled up or down as needed.
%Hypothesis generation coerces disparate computational results into a digestible form which can be easily parsed into the Analysis dataclass.
%The Analysis dataclass is a high-level summary of the results for a given agent that balances the Reasoning agent context limit with a rich representation of how successful and in what ways an agent successfully contributed to the hypothesis.
%Additionally, the Analysis object provides a clean way to perform checkpointing as it provides a history of agentic contributions for restarting runs, as well as for distributing runs across disconnected HPC clusters, workstations, etc.

\subsubsection{Managing soft exceptions}
For many computational tools in the broader domain of biology, the existence of soft exceptions remains a barrier to adoption by the layperson.
Similarly, we must consider such failure modes when building an agentic framework.
A clear, and abundant example is when MD simulations crash, often not communicating the error properly beyond messages such as ``Particle coordinates NaN'' or the opaque ``Segmentation fault.''
These types of crashes can arise from several sources: poor starting configuration, poor atomic parameters, insufficient energy minimization, failures in upstream system building, and simulation engine instability.
While these can sometimes resolved simply by restarting a simulation, this is not always the case, and the diagnosis of such errors is not trivial, even for the expert user.
For this reason, we build several quality assurance agents that can be invoked by the Reasoning agent to explore why a tool has failed, and annotate its findings in the hypothesis so that the Reasoning agent can make an informed choice of how to proceed.
By investigating and annotating soft exceptions and errors, the Reasoning agent is empowered to choose whether to restart a task or shift its attention to other agents and tasks.

For the MD simulation agent, this involves confirming that the input structure has non-overlapping coordinates, non-zero coordinates, and probing the pairwise forces on each pair of atoms.
Although the latter can be a somewhat expensive operation, it often determines whether a potentially valuable result can be salvaged instead of being discarded.
Additionally, the Reasoning agent controls the deployment of this resource-intensive investigation, activating it only when its necessity is clearly indicated.

For other agents, such as the structure prediction or binder design agents, this follows a similar protocol.
The initial quality control process involves examining structure files for severe atomic clashes, improperly formatted outputs, and incompleteness in scoring data. 
A common, unexpected failure mode occurs during the conversion from CIF to PDB format, which is often required because many existing computational biophysics tools cannot ingest CIF files directly. 
This conversion may appear successful but can silently write corrupt lines, leading to subsequent downstream crashes.
\section{StructBioReasoner Evaluation}

\begin{figure*}[h!]
    \centering
    \includegraphics[width=0.95\linewidth]{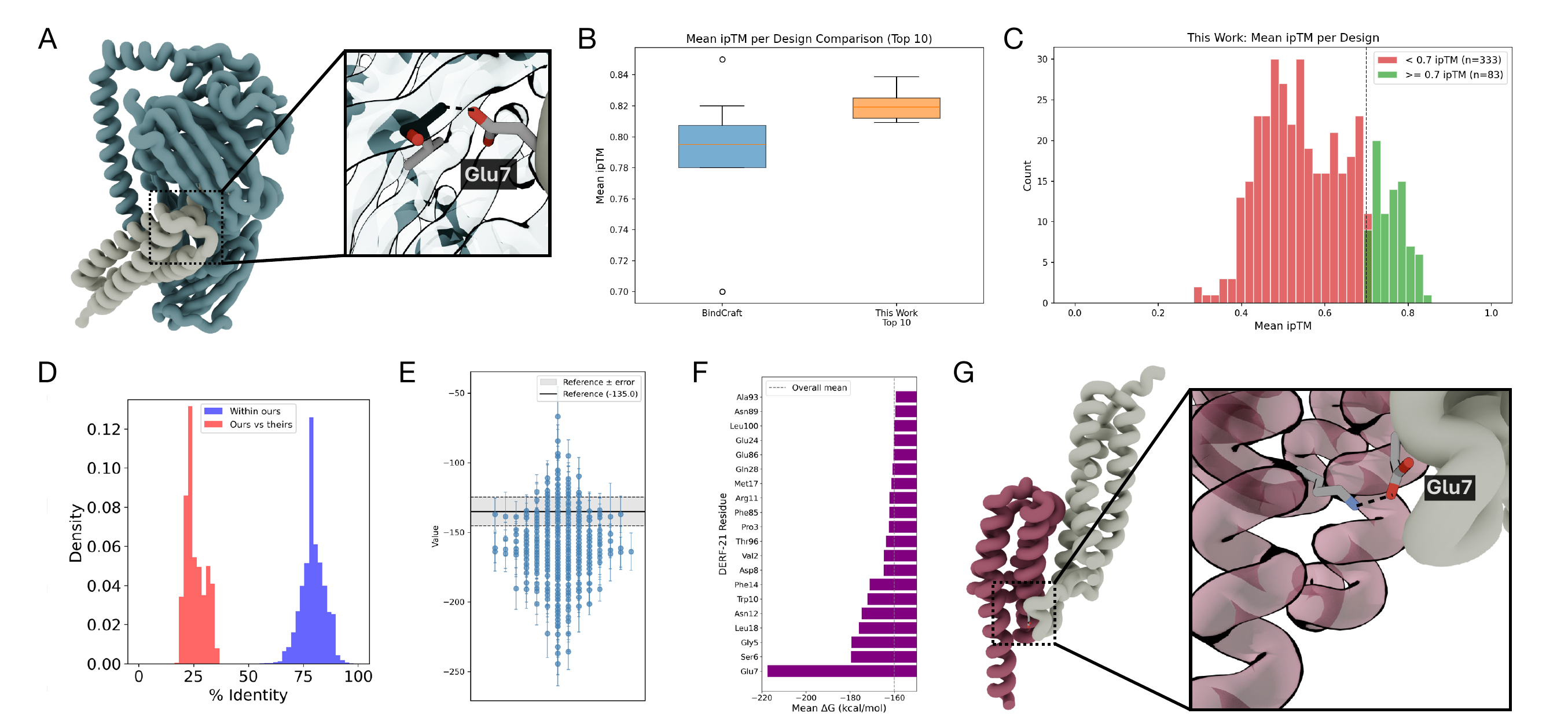}
    \caption{Evaluation of StructBioReasoner against Der f 21. 
    (A) Interactome simulation identified druggable interface in the IgE:Der f21 immunocomplex. Highlighted in the zoomed in view is glutamate residue 7, which forms a salt bridge with IgE (Derf21, and IgE colored in gray and blue respectively). 
    (B) \rev{Density plot of mean ipTM scores for each sequence of top 10 binders from this work compared to the 10 original BindCraft paper binders. Mean ipTM is higher with smaller variance in this work, while the best binder from BindCraft does have a slightly higher ipTM (0.86).}
    (C) \rev{Histogram of all mean ipTM scores for all designed binders. In green are all models with ipTM greater than 0.7, signifying the worst previously known binder. In this work 20\% of binders fall within this cutoff for a total of 83 total binders.}
    (D) Percent sequence identity was computed using the Needleman-Wunsch algorithm, within our designs (blue) and between our designs and BindCraft designs (red). Most cross-method comparisons fall below 30\%, indicating our binders are novel relative to prior designs.
    (E) Free energies shown in swarm plot. Reference binder energy shown in black with standard deviation shaded. (F) Average free energy of binding for binders which form each high frequency contact during simulation. High frequency contacts are defined as the top 20 most resident interactions within a distance cutoff of 3.0 \AA{}. (G) Molecular interface formed by top binder. Zoomed inset highlights targeting of E7 by a binder lysine, forming a salt bridge (Derf21 and binder colored in gray and red, respectively).}
    \label{fig:evaluation}
\end{figure*}

To validate StructBioReasoner, we designed binders for two benchmark systems:
(1) house dust mite allergen Der f 21 and
(2) Nicotinamide mononucleotide adenylyltransferase 2 (NMNAT-2).

\subsection{Der f 21 Binder Refinement}
Der f 21 provides a compact system with known epitopes~\cite{pang2019crystal} (highlighted in ~\autoref{fig:evaluation}A) and an experimentally validated protein binder designed via the BindCraft pipeline~\cite{pacesa2025one} which engages a distinct epitope.
Being a dust mite allergen, its small size and stable structure make it a particularly attractive testing target from a design and engineering perspective.
Additionally, in previous work~\cite{pacesa2025one}, several designed binders demonstrated activity with the best (binder 10) exhibiting a binding affinity of 793 nM as determined by surface plasmon resonance.
As a baseline reference, we performed 12 replicate simulations of Der f 21 in complex with binder 10, totaling 600 ns in aggregate and the binding free energy was calculated to be -135.00 $\pm$ 10.25 kcal/mol with MM-PBSA.

We then prompted the StructBioReasoner framework to design biologics binders for Der f 21.
We curated a vector store of papers freely available from PubMed to enrich the design cycles and let the Reasoning agent execute the design.
After two cycles of design by the Binder Design agent, we accumulated 842 total binders, of which 787 binders (93.47\%) passed quality control, structural verification and analysis checkpoints.
%This resulted in an \textit{in silico} design success rate of 93.47\%, meaning the ratio of designs, which passed both sequence and structure quality control.
Notably, in the designing process, the Reasoning agent mirrored human expert judgment, by relaxing the sequence quality control specially for designs of the literature peptide as the input sequence fails with the default settings for having too many charged residues and the His-tag exceeds the default number of same residue in a row.
A comparison was made to binders designed in previous work~\cite{pacesa2025one}, with the top 10 designed binders from StructBioReasoner demonstrating a higher mean ipTM score, with a lower variance than the top 10 literature binders.
The entire pool of successful designs was subjected to MD simulation and subsequent characterization by MM-PBSA calculations by the MD and Free energy agents, respectively.
These analyses yielded a 50.98\% success rate, wherein binders were found to have a more favorable free energy of binding than the top design from the compared work (\autoref{fig:evaluation}B-C).
We define a result as `more favorable' if its mean calculated free energy lies at least 1 standard deviation below -145.25 kcal/mol, which is a standard deviation from the mean free energy of the reference binder.

A previous study~\cite{pang2019crystal} identified the epitopes responsible for binding human IgE through mutagenesis experiments (K10, K26, K42, E43, K46, and K48).
However, the mutagenesis results from this study indicate moderate evidence that E7 plays a role in binding IgE and the E7A mutation abrogates severe symptoms in a majority of mice.
\rev{
Sequence identity analysis (~\autoref{fig:evaluation}D) demonstrates the novelty of our binder sequences compared to those from the previous study~\cite{pacesa2025one}. 
The large majority of cross-method sequence comparisons fall below the 30\% identity threshold, indicating that our designs are sequence-level novel relative to previously published binders for this target~\cite{brenner1998assessing}.
}
Designed binders frequently contact epitope-proximal residues such as E24 and F10 (~\autoref{fig:evaluation}F-G), with the caveat that our construct lacks the N-terminus from the previous work (i.e. K10 is K9 and K26 is K25).
This result, taken together with the dramatic improvements in binding free energy (~\autoref{fig:evaluation}E), serves to bolster support for the increasing use of agentic frameworks to supplant existing workflow-type experiments, enabling rapid and intelligent hypothesis generation, experiment testing and validation. 

\subsection{NMNAT-2 Biologics}
\begin{figure*}
    \centering
    \includegraphics[width=0.9\linewidth]{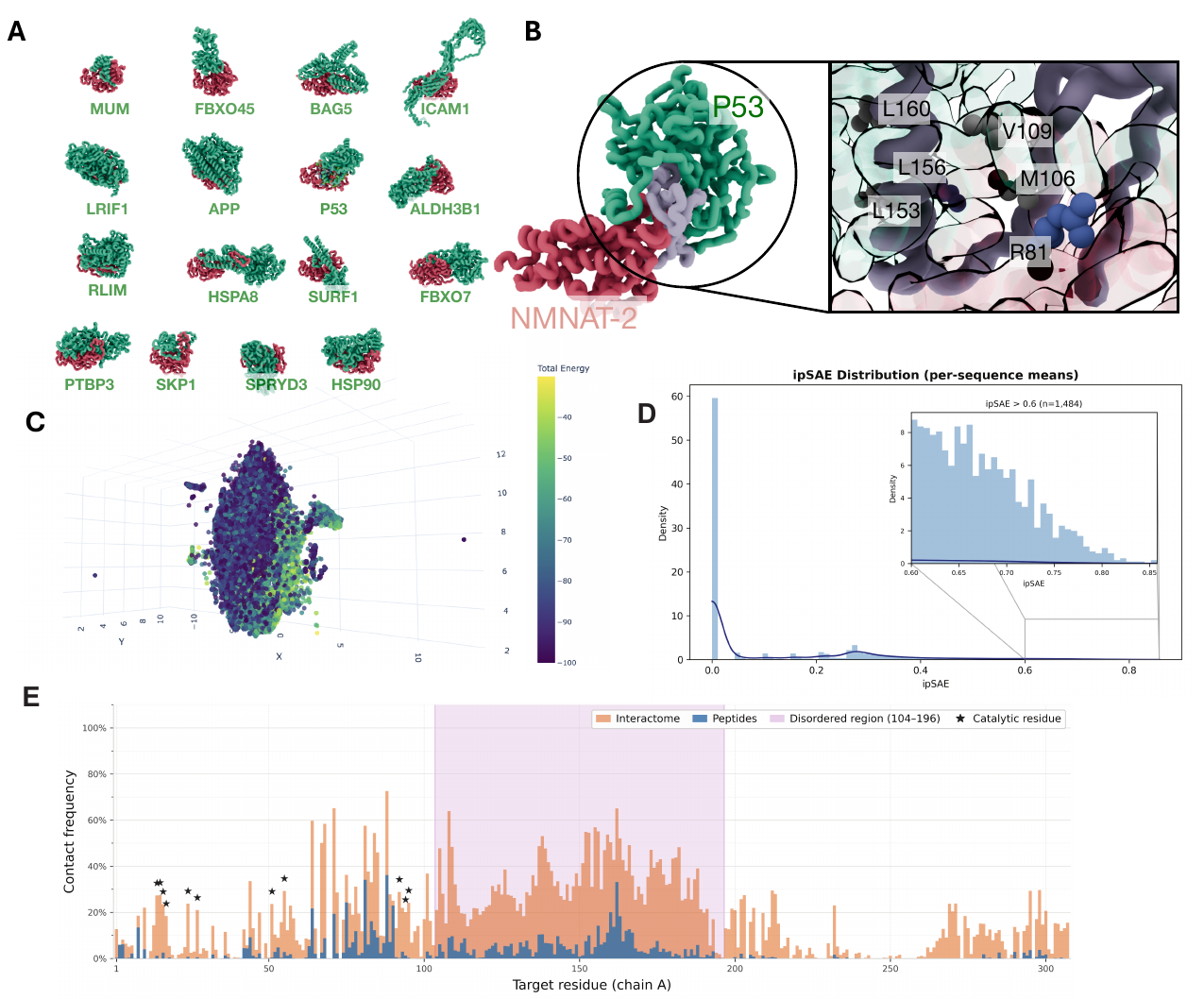}
    \caption{Evaluation of StructBioReasoner against NMNAT-2. Each panel represents a snapshot of what occurs during the various design and exploration tasks of the agentic framework. (A) Probing the NMNAT-2 interactome guided by RAG based reasoning. All interactants are colored in green while NMNAT-2 is red(B) The NMNAT-2:p53 interface serves as a potential binding site for biologics (NMNAT-2 disordered region colored in purple). (C) Protein embedding space for designed binders colored by electrostatic interaction energy; embedding space is represented with 3 components of UMAP dimensionality reduction. (D) \rev{Mean ipSAE scores for designed binders. 2.41\% of designs fall above a mean ipSAE of 0.6, indicating confidence in predicted interface.} \del{Refinement of binders by inverse folding, completing the binder design loop.} (E) Residue-wise contact map for all NMNAT-2/binder and NMNAT-2/interactant complex simulations. Contact frequency per target residue, defined as the proportion of interacting partners with at least one heavy atom within 3.0 Å of any heavy atom of that residue. Gray shaded area indicates the intrinsically-disordered region of NMNAT-2 (resid 104-196). Stars indicate residues in NMNAT-2's catalytic domain.}
    \label{fig:nmnat2}
\end{figure*}

NMNAT-2 is an important cancer-related protein with intrinsically disordered regions and a large number of interacting partners that requires a workflow to reason about specificity, interface disruption, and functional consequences—not just affinity.
%to generate biologics to disrupt the cancer target, NMNAT-2.
Owing to the difficulty of such an undertaking, we consider this a good benchmark to demonstrate the agentic approach in extending current co-scientist approaches that fall short of true experimentation. We prompted the framework to design biologics for NMNAT-2 and to augment its search with HiPerRAG \cite{gokdemir2025hiperrag}, leveraging our internal vector store of literature.
The framework went through a series of experiments wherein it first explored potential binding sites and pathway specificity by simulating 18 key members of the NMNAT-2 interactome identified via RAG.
By probing these interactome pairs, the Reasoner was able to identify multiple interacting partners such as Cadherin-related family member 3, DENN domain-containing protein 2D, and importantly, the ubiquitous interacting partner p53.
The NMNAT-2:p53 interface was identified as a target, hitting not only NMNAT-2 but also the well-studied cancer target p53.

After execution of the interactome simulations by the MD agent, the Reasoner analyzed interfaces between NMNAT-2 and each interacting partner (\autoref{fig:nmnat2}A).
The interactome interface deduced by the reasoner is predominantly mediated by IDR interactions, highlighting the importance of targeting this region.
This analysis identified a hydrophobic interface between p53 and the intrinsically disordered region of NMNAT-2 which it hypothesized can be disrupted to interfere with biologically-relevant downstream signaling pathways (\autoref{fig:nmnat2}B).
With this interface chosen, the framework then proceeded to design binders using a combination of the Binder Design, MD, and Free Energy agents.
This involved a series of rounds of design wherein an arbitrary starting scaffold miniprotein binder was co-folded against the target, and several rounds of inverse folding were performed to refine the resulting predicted binders. 
\rev{Co-folded designs were evaluated by ipSAE using a PAE cutoff of 12\AA, chosen as a sane middle ground between 10-15\AA\ as benchmarked in the ipSAE paper~\cite{dunbrack2025res}.
The fraction of designs with an ipSAE greater than 0.6 are highlighted in \autoref{fig:nmnat2}D, again chosen based on benchmarks in the source material as a reasonable cutoff for a higher confidence interface~\cite{dunbrack2025res}.}
It is illustrative to note that the miniprotein binder designs stabilize a set of electrostatic interactions that cluster distinctly in the embedding space (as determined by the evolutionary scale model/ ESM-2 650M) compared to unstable binders\rev{, shown in \autoref{fig:nmnat2}C.} 
\del{Further, as highlighted in \autoref{fig:nmnat2}D, a small number of residue in NMNAT-2 disordered region mediate the minibinder interactions.}
After generating 97,066/266,606 binders against NMNAT-2 that passed both sequence quality control and structural validation, each was subjected to MD simulation.

Binders were simulated in explicit solvent for 10 ns in atomistic detail, and the resulting trajectories were analyzed for stability by RMSD and RMSF calculations.
Finally, to better characterize each binder, the Reasoner computed the pairwise linear interaction energy between target and binder.
This analysis elucidated three major binding modes among the successful binders, one of which corresponds to the NMNAT2:p53 interface.
\rev{
Two of these three major binding modes target the IDR of NMNAT2. 
\autoref{fig:nmnat2}E shows the relative frequency of binder contacts per NMNAT-2 residue, demonstrating that our workflow designs binders for both IDR and non-IDR regions, which maps onto the interaction pattern in the native interactome. 
84.5\% of all binders that pass sequence quality control and sequence validation target at least one residue in the IDR of NMNAT-2. Notably, interfaces involving non-IDR residues are also observed in native interactome complexes, supporting their biological relevance (\autoref{fig:nmnat2}E). 
}

\section{StructBioReasoner Scaling and Performance}
To assess the scaling of our framework, we first benchmarked the performance of each individual agent.
For each agent we measured the appropriate metric of throughput: µs aggregate sampling per hour for the MD agent, binders/hour for the Binder Design agent, and MM-PBSA calculations per hour for the Free Energy agent.
%We also measured scaling on from 1--256 nodes.
%For reference, a single node on the ALCF Aurora supercomputer has 12 XPU accelerators, allowing us to leverage at most 3072 accelerators for these benchmarks.

We performed our set of baseline benchmark experiments on 64 nodes to establish a basis from which the ideal theoretical scaling can be calculated. 
We chose 64 nodes as our baseline, as it demonstrates substantially higher stability and reproducibility compared to the traditional baseline node count of 1. Benchmarks employing smaller node counts are also shown, but demonstrate inconsistent timing behavior, requiring a large quantity of benchmark instances to properly capture the fluctuations in scaling performance.

\subsection{Hardware Platforms}
Aurora \cite{allen2025aurora, ibeid2025scaling} is an HPE Cray EX system comprising 10,624 compute nodes interconnected by HPE Slingshot-11 in a Dragonfly topology. 
Each compute node consists of two Intel Sapphire Rapids Xeon CPU Max Series processors with 52 physical cores each (two hardware threads per core), each with 64 GB of on-package HBM and 512 GB of DDR5 memory, and six Intel Data Center GPU Max Series (Ponte Vecchio) XPUs. 
Each XPU has two tiles that can effectively act as separate accelerators, for a total of 12 tiles per node; tiles have 64 GB of high-bandwidth memory (HBM2e). 
The XPUs within a node are connected via Intel Xe Link interfaces in an all-to-all topology. The nodes are connected with 8× HPE Slingshot-11 NICs, providing a node-injection bandwidth of 200 GB/s. 
Each Ponte Vecchio tile is capable of delivering a peak of 17 TFLOPS in FP64 and 23 TFLOPS in FP32, for a full-node peak performance of 187 TFLOPS in FP64 and 267 TFLOPS in FP32.

\subsection{HiPerRAG Agent Scaling}

The scaling characteristics of the RAG framework employed in this work have been extensively documented in a previous study~\cite{gokdemir2025hiperrag}. 
That work demonstrated strong scaling performance across distributed platforms, scaling up to 1600 accelerators on Polaris, Sunspot, and Frontier supercomputers. 
Given that our current implementation utilizes HiPerRAG without architectural modifications, we refer readers to our prior publication for comprehensive scaling data and performance analysis.

%\subsection{Structure prediction agent scaling}

\subsection{MD Agent Scaling}

\begin{figure*}[h!]
    \centering
    \includegraphics[width=0.95\linewidth]{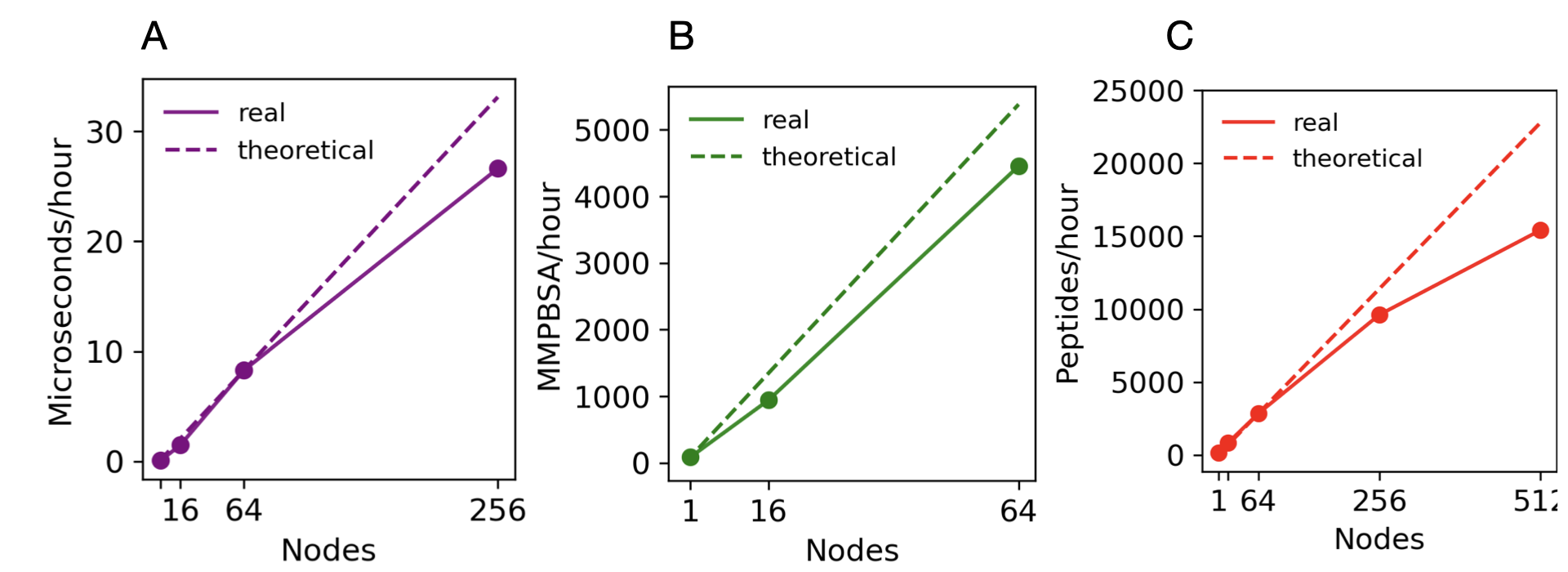}
    \caption{Scaling individual StructBioReasoner agents on Aurora. 
    (A) MD Simulation Agent scaling up to 256 nodes utilizing 3072 XPU accelerators, reported as aggregate simulation time per hour. 
    (B) Free Energy Agent scaling up to 64 nodes utilizing 12,800 CPU cores.
    (C) Binder Design Agent scaling up to 512 nodes utilizing 6144 XPU accelerators, reported as total number of designed peptides per hour.
    %, reported as . 
    %Demonstrates good weak scaling for experiments from 1--256 nodes, yielding upwards of 25 microseconds/hour of simulation at 256 nodes. Bottom: Scaling of the BindCraft binder design agent on Aurora, reported as total number of designed peptides per hour. Demonstrates good weak scaling for experiments from 1--512 nodes, with a throughput of over 15,000 peptides/hour at 512 nodes.
    }
    \label{fig:combined_scaling}
\end{figure*}

Simulations were performed by the MD Agent on target-binder systems generated by the Binder Design Agent. 
~\autoref{fig:combined_scaling}A shows the aggregate throughput of implicit-solvent MD simulations that contain approximately 5000-6000 atoms per system. 

%Explicit-solvent MD simulation benchmarks of the same systems (not shown) typically show around half the throughput
The MD simulation agent demonstrated robust weak scaling up to 256 nodes, yielding an aggregate sampling rate of around 26.6 µs/hour or 638.4 µs/day. 
It achieves 80.4\% efficiency on 256 nodes compared to the baseline run at 64 nodes. 

The MM-PBSA analysis agent showed weak scaling up to 64 nodes, processing MD trajectory outputs with 82.8\% parallel efficiency at the max node count tested as demonstrated in~\autoref{fig:combined_scaling}B. 
Beyond this regime, the agent became severely I/O-bound due to the high volume of trajectory data read from the parallel filesystem for chunking and post-processing analysis. 
A run on 256 nodes revealed that filesystem contention and bandwidth saturation dominate the runtime, negating computational gains from increased parallelism. 
Consequently, we limited MM-PBSA operations to at most 64 nodes in production workflows, where the balance between computational throughput and I/O overhead remained favorable.
For these reasons, we chose the single node benchmark as our baseline for theoretical scaling, in contrast to our other agents. 
Future work will explore staging strategies and in-memory data pipelines to mitigate filesystem bottlenecks at higher concurrency levels.

However, this constraint did not impact overall pipeline throughput, as the MM-PBSA agent completed over 4000 calculations per hour at 64 nodes, which was sufficient to evaluate all top candidates filtered from the $\sim$15,000 peptides generated hourly by the design agent. 
The multi-stage filtering architecture, which applied cheaper interaction energy calculations before MM-PBSA evaluation, ensured that only the most promising binders reached the computationally expensive MM-PBSA stage, making the 64-node limit a non-bottleneck in practice. \vspace{-0.2in}

%The MD simulations performed within our workflow are in principle embarrassingly parallel. 
%Thus, we might expect to reach closer to the ideal theoretical scaling at higher node counts. 
%We (predict/speculate/believe) that this slowdown is caused by low-level I/O operations moving files between local node memory (/dev/shm/) and disk. 

%\begin{figure}[h!]
%    \centering
%    \includegraphics[width=\linewidth]{FIGS/bc_scaling.png}
%    \caption{Scaling of the BindCraft binder design agent on Aurora. Demonstrates good weak scaling for experiments from 1 to 256 nodes, with a throughput of over 10,000 peptides/ hour.}
%    \label{fig:bcscaling}
%\end{figure}

\subsection{Binder Design Agent Scaling}
%Our framework enables two major computational workflows for peptide design: BindCraft~\cite{pacesa2025one} and Chroma~\cite{ingraham2023illuminating}.
%
% Chroma scales differently depending on the number of amino acids in the protein structure being diffused. We performed scaling benchmarks on proteins with varying length and achieve good weak scaling across all protein sizes tested
% Chroma performs backbone diffusion and inverse folding natively, while our bindcraft implementation relies on co-folding (structure prediction agent section 3.3) to generate structures and ProteinMPNN to generate sequences. The additional overhead introduced by low-level communication in BindCraft results in lower overall throughput, and remains a focus for optimization of the overall workflow.

%Figure~\ref{fig:combined_scaling}b 
\autoref{fig:combined_scaling}C shows the aggregate throughput of BindCraft in peptides per hour. The BindCraft agent demonstrated good weak scaling, achieving 84.4\% efficiency at 256 nodes and 67.7\% efficiency at 512 nodes compared to the baseline benchmark at 64 nodes. 

Given that the two major tasks performed within our BindCraft agent (forward folding and inverse folding) are embarrassingly parallel, we anticipated near-ideal scaling with node count. 
However, we observed that I/O operations, specifically file transfers between node-local memory (/dev/shm/) and the Aurora filesystem, which introduced a bottleneck that degrades efficiency beyond 256 nodes. 
This I/O overhead scaled superlinearly with node count as the aggregate filesystem load increased, limiting the overall parallel efficiency despite individual BindCraft tasks being compute-bound.

%\begin{figure}[h!]
%    \centering
%    \includegraphics[width=\linewidth]{FIGS/Diffusion_scaling.png}
%    \caption{Scaling of the Chroma diffusion agent on Aurora. Demonstrates good weak scaling across resource and task size, yielding upwards of 200 million peptides/day for NMNAT2.}
%    \label{fig:chroma}
%\end{figure}

%\subsection{Free Energy agent scaling}

%\subsection{DPO scaling}
%The scaling characteristics of the preference optimization framework employed in this
%work have been extensively documented in a previous study~\cite{dharuman2024mprot}.
%That work demonstrated strong scaling performance across distributed platforms, scaling up to 3000 nodes on Aurora achieving 4 exaFLOPS scaling for finetuning a 3.5 billion parameter protein language model.
%Given that our current implementa-
%tion utilizes HiPerRag without architectural modifications, 
%We refer readers to our prior publication for comprehensive scaling data and
%performance analysis.
%\begin{figure}[h!]
%    \centering
%    \includegraphics[width=\linewidth,trim=13mm 4mm 8mm 16mm,clip]{FIGS/DPO_scaling.png}
%    \caption{Preference optimization scaling for 3.5B parameter model on different compute resources. Each tested system is capable of breaking the exascale barrier for this task.}
%    \label{fig:dpo}
%\end{figure}

\section{Conclusion}
We presented StructBioReasoner, a scalable multi-agent framework for autonomous biologics design targeting intrinsically disordered proteins.
By integrating retrieval-augmented generation, structural prediction, molecular dynamics simulation, and iterative binder design within a unified reasoning architecture, our system addresses the unique challenges posed by IDP targets:
specifically, their conformational heterogeneity, the fragmentation of existing computational tools and workflows, and the substantial computational resources required for exploration of the design space.
The tournament framework enables parallel hypothesis generation and evaluation, while the cross-hypothesis learning mechanism allows the system to accumulate target-class-specific design intuition across campaigns without explicit model retraining.
We evaluated on two benchmark systems demonstrates the practical utility of this approach.
%\del{For Der f 21, the framework achieved a 93.47\% \textit{in silico} design success rate and identified binders with substantially improved binding free energies compared to previously validated designs, with over 50\% of candidates outperforming the reference binder.}
%For Der f 21, the system autonomously identified E7 as a key interaction residue and designed binders against this, corroborating mutagenesis data from the literature.
%For the more challenging NMNAT-2 target, the framework navigated a complex interactome to identify the NMNAT-2:p53 interface as a therapeutically relevant binding site and generated diverse binder candidates. 
Our results suggest that agentic reasoning can match or exceed human-guided workflows in identifying biologically meaningful design strategies for IDP targets.
While these in silico results are encouraging, we recognize that experimental validation remains an important next step to fully establish the therapeutic viability of the designed binders. 
As with any computational screening approach, in silico performance may not directly translate to experimental success, and we view these findings as strong computational hypotheses that motivate promising future wet-lab characterization. 
%To this end, future ablation studies will play an important role in characterizing the exact role of agentics in this design pipeline.
Rigorous ablation studies are planned as follow-on work, systematically comparing RAG-grounded versus vanilla LLM reasoning, single-pass versus tournament-based design, and DPO-optimized versus unoptimized candidate generation to isolate the contribution of each component to overall workflow performance.
Component-level signals within the existing results provide implicit evidence of each layer's functional contribution. This includes (1) HiPerRAG-grounded interactome simulation step that provides a principled basis for identifying the NMNAT-2:p53 interface as a therapeutically relevant target, (2) comparison against BindCraft binder 10 (the best single-shot human-guided design from the prior literature) provides a natural baseline against which the agentic loop can be evaluated, and sequence identity analysis (Figure 3D) demonstrates that the vast majority of our designs fall below 30\% identity relative to all known BindCraft binders for Der f 21, confirming that the framework is exploring genuinely novel sequence space rather than recapitulating prior solutions. 
A structured factorial comparison across these axes will be reported in a dedicated follow-on study; however, these studies also expose the intensive computational workflows needed to establish these benchmarks.
These results nonetheless suggest that agentic reasoning can match or exceed human-guided workflows in identifying biologically meaningful design strategies for intrinsically disordered targets.

%From a computational perspective, the framework demonstrates strong scaling characteristics on the Aurora supercomputer, with the MD simulation agent achieving 80.4\% efficiency at 256 nodes and the binder design agent maintaining 84.4\% efficiency at the same scale. 
%The identification of I/O bottlenecks beyond 256 nodes for certain agents provides clear direction for future optimization, including the development of staging strategies and in-memory data pipelines. 
Our work establishes a foundation for deploying autonomous scientific reasoning at exascale, where the combination of vast computational resources and intelligent agent orchestration can accelerate the discovery of therapeutics against historically intractable targets.
Addressing the identified scaling challenges—particularly through distributed reasoning architectures, I/O optimization, and formal multi-agent coordination protocols—represents a clear path toward robust exascale deployment. 
Future work should systematically characterize agent interaction dynamics, develop metrics for collaborative efficiency, and explore hierarchical or federated reasoning architectures that can maintain strategic coherence while fully exploiting the parallelism available in next-generation supercomputers. 
As HPC resources continue to scale and AI methods for structural biology mature, frameworks like StructBioReasoner will play an increasingly central role in translating computational advances into therapeutic impact.
\rev{ This framework naturally extends to multiple parallel Reasoning Agents each targeting distinct hotspots, representing a promising direction for future work.}

\bibliographystyle{ACM-Reference-Format}
\bibliography{bibfile}

\end{document}